\documentclass[onecolumn,aps,prc,superscriptaddress,preprintnumbers,amssymb,floatfix,nofootinbib]{revtex4}
\usepackage{epsfig,graphics}
\usepackage{graphicx}
\usepackage{dcolumn}
\usepackage{bm}
\usepackage{amsmath}
\usepackage[usenames]{color}
\usepackage{ulem} 
\usepackage{booktabs}
\usepackage{array}  
\usepackage{CJK}
\usepackage[colorlinks,linkcolor=blue,urlcolor=blue,citecolor=blue]{hyperref}
\usepackage{hyperref}
\usepackage{tikz}
\usepackage{csquotes}
\usetikzlibrary {shapes.geometric}
\usetikzlibrary{shapes,arrows}
\begin{document}
\begin{CJK*}{UTF8}{gbsn}

\title{Bayesian extraction of TMC-free collectivity in proton{\textemdash}proton and proton{\textemdash}nucleus collisions at the LHC}

\author{Shuang Guo}
\email[]{guos19@fudan.edu.cn}
\affiliation{Key Laboratory of Nuclear Physics and Ion-beam Application (MOE), Institute of Modern Physics, Fudan University, Shanghai 200433, China}
\affiliation{Shanghai Research Center for Theoretical Nuclear Physics, NSFC and Fudan University, Shanghai $200438$, China}

\author{Jia-Lin Pei}
\email[]{jlpei22@m.fudan.edu.cn}
\affiliation{Key Laboratory of Nuclear Physics and Ion-beam Application (MOE), Institute of Modern Physics, Fudan University, Shanghai 200433, China}
\affiliation{Shanghai Research Center for Theoretical Nuclear Physics, NSFC and Fudan University, Shanghai $200438$, China}

\author{Guo-Liang Ma}
\email[]{glma@fudan.edu.cn}
\affiliation{Key Laboratory of Nuclear Physics and Ion-beam Application (MOE), Institute of Modern Physics, Fudan University, Shanghai 200433, China}
\affiliation{Shanghai Research Center for Theoretical Nuclear Physics, NSFC and Fudan University, Shanghai $200438$, China}

\author{Adam Bzdak}
\email[]{bzdak@fis.agh.edu.pl}
\affiliation{AGH University of Krakow,\\
Faculty of Physics and Applied Computer Science,
30-059 Krak\'ow, Poland}

\begin{abstract}
A central challenge in understanding the origin of collective flow-like signatures in small collision systems calls for a reliable method to disentangle collective flow from substantial background correlations, especially those arising from transverse momentum conservation (TMC). A Bayesian inference framework is developed to integrate TMC calculations with the LHC-ATLAS data on long-range multiparticle azimuthal correlation observables, thereby extracting TMC-free collectivity in small systems. Our analysis indicates that while the TMC-free elliptic and triangular flow ($v_{2}$ and $v_{3}$) are similar, the $p$+$p$ and $p$+Pb systems exhibit distinct TMC backgrounds, TMC-flow interplay, and $v_{2}$--$v_{3}$ correlations. We demonstrate that the extracted $v_{2}$ and $v_{3}$ are well described by the measured four-particle $v_2\{4\}$ and two-particle $v_3\{2\}$ in $p$+Pb collisions, whereas these measurements systematically underestimate the TMC-free flow in $p$+$p$ collisions, due to competing contributions from TMC effects. This establishes a robust and data-driven framework, enabling a quantitatively controlled interpretation of collective signals and opening a new avenue for understanding the origin of collectivity in small collision systems.
\end{abstract}


\maketitle

\vspace{0.5em} 

\section{INTRODUCTION}
\label{introduction}

Quark-gluon plasma (QGP), a deconfined state of matter where quarks and gluons are liberated from confinement, forms in the extreme, hot and dense environment created by high-energy nucleus-nucleus collisions at the Relativistic Heavy Ion Collider (RHIC) and the Large Hadron Collider (LHC) \cite{qgp1,qgp2,qgp3,qgp4,qgp5,qgp6,NST_Chen1,NST_Shou,Duan:2025wsy,Ma:2026nst}. The experimental results have shown that this new type of nearly perfect fluid can translate initial spatial geometry or initial energy density fluctuations into momentum anisotropy of the final particles through the pressure gradient in hydrodynamics \cite{hydro2,hydro10,hydro7,hydro8,hydro1,hydro5,hydro9,hydro3,hydro6,hydro4,NST_Jia}, resulting in the emergence of a strong collective flow in A+A collisions \cite{hydro11,hydro12, NST_Ma2}. To quantify azimuthal anisotropy in the transverse momentum plane, the particle distribution is parameterized by a Fourier series $\frac{dN}{d\varphi} \propto 1 + \sum_{n} v_n \cos[n(\varphi - \Psi_n)]$. This expansion defines the anisotropic flow coefficients, where $v_1$, $v_2$, $v_3$, and $v_4$ correspond to directed, elliptic, triangular, and quadrangular flow, respectively \cite{Four2,Four1,Four3}.  Experimentally, flow coefficients are extracted using various methods, including the event plane method \cite{event_plane2,event_plane1}, two-particle correlations and multi-particle cumulants $v_n\{k\}$ \cite{two_particle1}. The latter is particularly valued for its ability to suppress non-flow effects \cite{cumulant1,cumulant2}. These non-flow correlations, which contaminate the flow signal, originate from various sources \cite{nonflow0}: short-range ones such as jets, resonance decays, and Bose-Einstein correlations, as well as long-range ones like those induced by transverse momentum conservation (TMC) \cite{non_flow3,non_flow2,non_flow1,non_flow4}.

Over the past decade, collective flow measurements have been performed in a variety of colliding systems, including $p$+$p$ and $p$+Pb collisions at the LHC \cite{LHC1,LHC2,LHC3,LHC4,LHC5,LHC6,Chen:2024eaq}, as well as $p$+Au, $d$+Au, and $^3$He$+$Au collisions at RHIC \cite{RHIC1,RHIC2,RHIC3,RHIC4,RHIC5}. Surprisingly, the ridge structure, characterized by a small azimuthal separation and an extended longitudinal correlation, which is regarded as direct evidence for collective flow in A+A collisions \cite{ridge_structure1,ridge_structure2}, has also been observed in small systems such as $p$+$p$ and $p$+A. This finding raises the question of whether QGP could also be produced in such small colliding systems. Numerous theoretical efforts have been made to explain the emergence of collective flow in small colliding systems. The proposed explanations can be broadly categorized into several mechanisms. Firstly, similar to the scenario in large systems, hydrodynamic models describe how the pressure gradient of a possible QGP droplet in small systems converts initial geometric asymmetries into final-state momentum anisotropy \cite{hydro_small1,hydro_small3,hydro_small4,hydro_small5,hydro_small2,hydro_small6,hydro_small7}. Alternatively, transport models attribute flow signals to a parton escape mechanism, where the anisotropic escape probability of partons generates the observed $v_n$ \cite{transport_small1,transport_small2,transport_small3,transport_small4,transport_small5_nst,Bay_new2}.

Conservation laws represent another significant source of azimuthal correlations among particles \cite{globalMC,conservation_law1,conservation_law2,conservation_law3,conservation_law4}, among which transverse momentum conservation (TMC) plays a crucial role in explaining collective flow behavior in small colliding systems. We have found that the TMC effect induces positive \(2k\)-particle elliptic flow cumulants \(c_2\{2k\}\), which scale as \(1/(N - 2k)^{2k}\) \cite{TMC1}, where $N$ is the number of particles constrained by transverse momentum conservation. Furthermore, when TMC interacts with hydrodynamics-like elliptic flow, the four-particle cumulant \(c_2\{4\}\) undergoes a sign change. This behavior provides a natural explanation for the observed multiplicity-dependent sign change of \(c_2\{4\}\), \(c_3\{2\}\) and \(c_3\{4\}\) in small systems at the LHC \cite{TMC2,TMC3}. Our recent calculations of the four-particle symmetric cumulants \(sc_{2,3}\{4\}\) and \(sc_{2,4}\{4\}\) \cite{symmetric} show that the observed correlations originate from TMC at low multiplicities, with collective flow becoming increasingly dominant as the particle number \(N\) rises, which aligns well with ATLAS measurements \cite{TMC4,TMC5}.

Since the TMC theoretical framework has demonstrated good agreement with experimental observations in small collision systems, it is natural to explore whether collective flow signals can be quantitatively extracted from experimental data within this approach. Our goal is to identify a set of collective flow parameters that can simultaneously describe multiple experimental observables in small systems. In previous studies, calculations of multi-particle correlators $c_n\{k\}$ that include both TMC and flow effects typically rely on manually fixing key parameters, such as the elliptic and triangular flow coefficients $v_2$ and $v_3$, as well as the transverse momentum of the correlated particles and the event-averaged $\langle p^2 \rangle$. While this tuning strategy can reproduce individual observables, different parameter choices are often required for different measurements, limiting the predictive power of the approach and preventing a systematic estimation of parameter uncertainties.

A more robust and principled approach to extract these parameters directly from experimental data is offered by Bayesian inference \cite{BAY0}. This statistical framework provides a powerful tool for inverse problems, allowing for the simultaneous determination of multiple parameters and a rigorous assessment of their uncertainties. Within heavy-ion physics, Bayesian methods have already been successfully employed for the precise extraction of fundamental properties, such as the specific shear viscosity \cite{BAY1,BAY2,BAY3,BAY4,BAY5,BAY6,BAY7,BAY8}, nuclear structure \cite{BAY11,BAY12}, the equation of state (EoS) of nuclear matter \cite{BAY9,BAY10,Bay_new3,Bay_new4,Bay_new1}, and jet quenching properties \cite{Bay_new5}, demonstrating their capability to constrain theoretical models with complex dependencies. Therefore, in this work, we will simultaneously constrain the {TMC-free collectivity} parameters $v_2$ and $v_3$ in $p$+$p$ and $p$+Pb collision systems using four experimental observables $c_2\{4\}$, $c_3\{2\}$, $c_3\{4\}$, and $sc_{2,3}\{4\}$ from the ATLAS experiment \cite{ATLAS_c32,ATLAS_c24c34,ATLAS_sc23}. In the present work, ``TMC-free collectivity" refers operationally to the flow components ($v_2$, $v_3$) obtained after subtracting the modeled TMC contributions. We further incorporate the dependence of collective flow on charged-particle multiplicity \(N_\text{ch}\) by parameterizing its functional form. This approach not only enables the unified extraction of collective flow properties from experimental measurements, but also facilitates the investigation of the collective-flow component after accounting for the modeled TMC contribution.

This paper is structured as follows. Section~\ref{TMC} outlines the calculation of cumulants, including \( c_2\{4\} \), \( c_3\{2\} \), \( c_3\{4\} \), and the four-particle symmetric cumulant \( sc_{2,3}\{4\} \), based on transverse momentum conservation and flow effects. Section~\ref{Bayesian} introduces Bayes' theorem and elaborates on the specific Bayesian inference procedure employed in this work. In Sec.~\ref{results}, we present results obtained under two distinct parameter settings, which provide an excellent description of the experimental data. In particular, the inferred collective flow coefficients $v_2$ and $v_3$ are compared with the ATLAS measurements of multi-particle azimuthal correlations using the sub-event cumulant method. We further provide a detailed discussion of the collective flow behavior in the two small collision systems, together with a comparative analysis of the different physical contributions in $p$+$p$ and $p$+Pb collisions. Finally, Sec.~\ref{summary} summarizes our findings and highlights their potential implications for investigating collective flow in small collision systems, with an emphasis on suppressing non-flow backgrounds.

\section{Cumulant and Symmetric Cumulant from TMC and flow}
\label{TMC}
We begin by outlining the theoretical framework for transverse momentum conservation (TMC), which serves as a baseline source of correlations among final-state particles. In an $N$-particle system with imposed transverse momentum conservation, i.e., $\vec{p }_{1}+\dots +\vec{p }_{N}=0$, the $k$-particle ($k< N$) transverse momentum probability distribution $f_{k}(\vec{p}_{1}, \dots, \vec{p}_{k})$ is given by~\cite{conservation_law4,TMC1,TMC2}:
\begin{equation}
\begin{split}
f_{k}(\vec{p}_{1},\dots,\vec{p}_{k}) = & f(\vec{p}_{1})\cdots f(\vec{p}_{k}) \frac{N}{N-k} \\
& \times \ \mathrm{exp}\left ( - \frac{\left ( p_{1,x}+ \dots   +p_{k,x} \right )^{2} }
{2\left (  N-k\right )\left \langle p_{x}^{2}  \right  \rangle _{F} }
  - \frac{\left ( p_{1,y}+ \dots   +p_{k,y} \right )^{2}  }
{2\left (  N-k\right )\left \langle p_{y}^{2} \right \rangle _{F} }\right ),
\label{fk}
\end{split}
\end{equation}
where $f(\vec{p})$ denotes the single-particle distribution function, with only $v_2$ and $v_3$ effects considered in our calculations,
\begin{equation}
    f\left ( \vec{p} \right )=\frac{g\left ( p \right ) }{2\pi}\left ( 1+2v_{2}\left ( p \right ) 
\cos \left [ 2\left ( \phi -\Psi _{2}\right )  \right ]+2v_{3}\left ( p \right )
\cos \left [ 3\left ( \phi -\Psi _{3}\right )  \right ]   \right ),
\end{equation}
\begin{equation}
p_{x} = p\cos \left (  \phi\right ),
\hspace{0.5cm}p_{y} = p\sin \left (  \phi\right ),
\end{equation}
\begin{equation}
\left \langle p_{x}^{2}   \right \rangle_{F}  = \frac{1}{2} \left \langle p^{2}  \right \rangle _{F}\left ( 1+  v_{2F}   \right ),
\hspace{0.5cm}\left \langle p_{y}^{2}   \right \rangle_{F}  = \frac{1}{2} \left \langle p^{2}  \right \rangle _{F}\left ( 1-  v_{2F}   \right ),
\end{equation}
\begin{equation}
\left \langle p ^{2}\right \rangle_{F} =
\frac{\int_{F} p ^{2}f(\vec{p})d^{2}\vec{p}}{\int_{F} f(\vec{p})d^{2}\vec{p}},
\end{equation}
\begin{equation} 
v_{2F}=   \frac{\int_{F} v_{2}\left ( p \right )g(p)\cos \left (  2\Psi _{2} \right )  p^{2}d^{2}p    }{\int_{F} g(p)p^{2}d^{2}p},
\label{eq:v2F_def}
\end{equation}
and $F$ denotes the full phase space.

We compute the following cumulants within this framework: the four-particle elliptic cumulant $c_2\{4\}$, the two- and four-particle triangular cumulants $c_3\{2\}$ and $c_3\{4\}$, and the four-particle symmetric cumulant $sc_{2,3}\{4\}$. \\

For example,
\begin{equation}
    c_{2}\left \{ 4 \right \}= \left \langle e^{i2\left ( \phi _{1}+\phi _{2} 
 -\phi _{3}-\phi _{4}  \right ) }  \right \rangle-2
\left \langle e^{i2\left ( \phi _{1}-\phi _{2}  \right ) }  \right \rangle^{2}.
\label{eq:c24}
\end{equation}
It should be noted that, in our analytical calculations, event-by-event fluctuations of $v_2$, $v_3$, and $\left[ p \right]$, as well as their correlation coefficient $\rho_n \left( v_n\{2\}^{2}, [p] \right)$, are not considered. Therefore, only single brackets are used. The detailed procedures for calculating the cumulants are summarized as follows:
\begin{align}
\left \langle e^{i2\left ( \phi _{1}-\phi _{2}  \right ) }  \right \rangle \big|_{p_1,p_2} &=\frac{\int_{0}^{2\pi}e^{i2\left ( \phi _{1}-\phi _{2}  \right ) }
f_{2}(\vec{p}_{1},\vec{p}_{2})d\phi _{1} d\phi _{2}}
{\int_{0}^{2\pi}f_{2}(\vec{p}_{1},\vec{p}_{2})d\phi _{1} d\phi _{2}}\notag \\
&\approx  A_{0}+A_{1}Y_{A}+\frac{1}{2} A_{2}Y_{A}^{2},
\label{eqA0}
\end{align}
where the form of $f_{2}(\vec{p}_{1},\vec{p}_{2})$ is taken from Eq.~\eqref{fk}. Assuming $p_{1}=p_{2}=p$, we obtain:
\begin{equation}
Y_{A}=-\frac{p^{2} }{(N-2) \left \langle p^{2}  \right \rangle _{F}},
\label{eq:YA}
\end{equation}
and
\begin{align}
A_{0} &=v_{2}^{2},\notag\\
A_{1} &=2 v_{2}^{2} + v_{3}^{2} - v_{2} v_{2F} \cos(2 \Psi_{2}) ,\notag\\
A_{2} &=1 + 6 v_{2}^{2} + \frac{{v_{2F}^{2}}}{2} + 3 v_{2}^{2} v_{2F}^{2} + 4 v_{3}^{2} + 2 v_{2F}^{2} v_{3}^{2}  - 8 v_{2} v_{2F} \cos(2 \Psi_{2}) + \frac{{1}}{{2}} v_{2}^{2} v_{2F}^{2} \cos(4 \Psi_{2}).
\label{eqA}
\end{align}
The results in Eq.~\eqref{eqA} stem from expanding the exponential term in Eq.~\eqref{fk}. As for the numerator in Eq.~\eqref{eqA0}, we expand it to the second order, i.e., $\text{exp} \left ( -X \right ) \approx 1-X+\frac{X^{2}}{2} $, since the first non-vanishing pure TMC term, the first term of $A_2$ in Eq.~\eqref{eqA}, appears at $\frac{X^{2}}{2}$. However, for the denominator in Eq.~\eqref{eqA0}, we expand it to the zero order, yielding a result of $\left ( 2 \pi \right ) ^{2} $. 
The physical interpretation of each term in the complex expressions of Eq.~\eqref{eqA} is summarized as follows:
\begin{itemize}
    \item Terms depending solely on $N$ and $p$ represent the pure TMC contribution. In this case, this term is given by $Y_{A}$.
    \item Terms depending only on $v_{n}$ and $\Psi_{n}$ originate from collective flow. This term is given by our $v_{2}^{2}$.
    \item Terms involving both $(N, p)$ and $(v_{n}, \Psi_{n})$ arise from the interplay between TMC and flow.
\end{itemize}
Later, we will discuss three distinct contributions to the observables in both $p$+$p$ and $p$+Pb collision systems, which are categorized as the ``TMC terms", the ``pure flow terms", and the ``interplay terms." \\
Similarly, we have 
 \begin{align}
        \left \langle e^{i2\left ( \phi _{1}+\phi _{2} -\phi _{3}-\phi _{4}  \right ) }  \right \rangle \big|_{p_1,p_2,p_3,p_4}&=\frac{\int_{0}^{2\pi}e^{i2\left ( \phi _{1}+\phi _{2} -\phi _{3}-\phi _{4}  \right )}
f_{4}(\vec{p}_{1},\cdots ,\vec{p}_{4})d\phi _{1} d\phi _{2}d\phi _{3}d\phi _{4}}
{\int_{0}^{2\pi}f_{4}(\vec{p}_{1},\cdots ,\vec{p}_{4})d\phi _{1} d\phi _{2}d\phi _{3}d\phi _{4}}\notag \\
        &\approx  B_{0} + B_{1}Y_{B} + \frac{1}{2} B_{2}Y_{B}^{2} +\frac{1}{6} B_{3}Y_{B}^{3} + \frac{1}{24} B_{4}Y_{B}^{4}
        \label{Bi}
\end{align}
where (assuming $p_{1}=p_{2}=p_{3}=p_{4}=p$)
\begin{equation}
Y _{B}=-\frac{p^{2} }{(N-4) \left \langle p^{2}  \right \rangle _{F}},
\label{eq:YB}
\end{equation}
and the full results for each order of $B$ in Eq.~\eqref{Bi} were presented in Ref.~\cite{TMC3}.

For $c_3\{2\}$, we have:
\begin{equation}
    c_{3}\left \{ 2 \right \}= 
\left \langle e^{i3\left ( \phi _{1}-\phi _{2}  \right ) }  \right \rangle,
\end{equation}
where
\begin{align}
\left \langle e^{i3\left ( \phi _{1}-\phi _{2}  \right ) }  \right \rangle
\big|_{p_1,p_2}&=\frac{\int_{0}^{2\pi}e^{i3\left ( \phi _{1}-\phi _{2}  \right ) }
f_{2}(\vec{p}_{1},\vec{p}_{2})d\phi _{1} d\phi _{2}}
{\int_{0}^{2\pi}f_{2}(\vec{p}_{1},\vec{p}_{2})d\phi _{1} d\phi _{2}}\notag \\
& \approx 
C_{0}+C_{1}Y_{C}+\frac{1}{2} C_{2}Y_{C}^{2}+\frac{1}{6} C_{3}Y_{C}^{3}.
\end{align}
Assuming $p_{1}=p_{2}=p$, we obtain
\begin{equation}
Y_{C}=-\frac{p^{2} }{(N-2) \left \langle p^{2}  \right \rangle _{F}},
\label{eq:YC}
\end{equation}
and
\begin{align}
C_{0} &=v_{3}^{2},\notag\\
C_{1} &=v_2^2 + 2 v_3^2 
,\notag\\
C_{2} &=4 v_2^2 + 2 v_2^2 v_{2F}^2 + 6 v_3^2 + 3 v_{2F}^2 v_3^2 
- 2 v_2 v_{2F} \cos(2 \Psi_2) 
,\notag\\
C_{3}& =1 + 15 v_2^2 + \frac{3 v_{2F}^2}{2} + \frac{45 v_2^2 v_{2F}^2}{2}
 + 20 v_3^2 + 30 v_{2F}^2 v_3^2 - 18 v_2 v_{2F} \cos(2 \Psi_2) 
- \frac{9}{2} v_2 v_{2F}^3 \cos(2 \Psi_2) 
\notag\\ &+ \frac{3}{2} v_2^2 v_{2F}^2 \cos(4 \Psi_2) - \frac{1}{4} v_{2F}^3 v_3^2 \cos(6 \Psi_3) .
\end{align}

In addition, for
\begin{equation}
    c_{3}\left \{ 4 \right \}= \left \langle e^{i3\left ( \phi _{1}+\phi _{2} 
 -\phi _{3}-\phi _{4}  \right ) }  \right \rangle-2
\left \langle e^{i3\left ( \phi _{1}-\phi _{2}  \right ) }  \right \rangle^{2}
\label{eq:c34}
\end{equation}
and
\begin{equation}
     sc_{2,3}\left \{ 4 \right \}=  \left \langle e^{i2\left ( \phi _{1}-\phi _{2}  \right )+i3\left ( \phi _{3}-\phi _{4}  \right ) }  \right \rangle - 
\left \langle e^{i2\left ( \phi _{1}-\phi _{2}  \right ) }  \right \rangle  \left \langle  e^{i3\left ( \phi _{3}-\phi _{4}  \right ) }  \right \rangle,
\label{eq:sc23}
\end{equation}
the specific results can be found in Refs.~\cite{TMC3,TMC5}.

The calculation involves several parameters, including the collective flow $v_n$, the transverse momentum  \( p \) of the selected \( k \)-correlated particles, the mean value of $p^2$ over the full space $\left \langle p^{2} \right \rangle_{F}$ and the value of $v_{2F}$, given by Eq.~\eqref{eq:v2F_def}. In principle, $N$, the number of particles subject to TMC, can be regarded as another parameter. However, as seen, e.g., in Eq.~(\ref{eq:YA}), it always appears together with $\left \langle p^{2} \right \rangle_{F}$. This issue will be discussed later. While these parameters are conventionally estimated using approximations from experimental data, we instead treat them as free parameters. Within a Bayesian framework, we infer the posterior distributions of these parameters directly from the experimental observables, thereby obtaining not only single-value estimates but also a complete characterization of their uncertainties. In addition, the event plane is also a parameter in the calculation results, but we find that the cumulants exhibit very weak dependence on it. Therefore, we treat it as a fixed value, where the value is taken from experimental measurements \cite{v2f_psi_set_exp}.

\section{Bayesian Inference}
\label{Bayesian}
Bayesian inference provides a probabilistic framework for progressively refining estimates of model parameters. It systematically updates prior beliefs with information from observed data to yield posterior distributions, thereby quantifying the uncertainty in parameter estimation. In heavy-ion physics, Bayesian inference has found widespread application in confronting theoretical models with experimental data to extract key model parameters. The fundamental principle of Bayesian inference lies in updating our beliefs about uncertain parameters in light of empirical evidence. This process is mathematically formalized through Bayes' theorem, which defines the posterior distribution of parameters $\theta$ conditioned on the observed data $D$ as, 
\begin{equation}\label{posterior}
 p(\theta|D)\propto p(D|\theta) p(\theta).
\end{equation}
In this probabilistic framework, each component plays a distinct role. The prior distribution $p(\theta)$ encapsulates our initial understanding or hypotheses about the parameters before observing the data. This can incorporate information from previous experiments, theoretical constraints, or physical boundaries, providing a rigorous means to integrate domain knowledge into the analysis. The likelihood function $p(D|\theta)$ serves as the core link between the computational model and experimental data. It quantifies the probability of observing the measured data $D$ under the assumption that the model with parameters $\theta$ is true, effectively measuring how well different parameter configurations explain the empirical observations. The posterior distribution $p(\theta|D)$ represents the complete solution to the inverse problem, synthesizing prior knowledge with information from the observed data. This distribution provides a complete probabilistic description of parameter uncertainties, offering not only point estimates but also credible intervals and correlations between parameters.

The central objective of this work is to determine the unconstrained parameters, specifically, the elliptic flow $v_2$, triangular flow $v_3$, the transverse momentum and $v_{2F}$, that characterize the cumulants derived from the combined effects of TMC and collective flow. We find that the theoretical predictions exhibit only a weak dependence on $v_{2F}$. Consequently, while $v_{2F}$ is included as a free parameter in the inference procedure, it does not play a central role in the subsequent analysis and will not be discussed further. In particular, as can be seen from Eqs.~(\ref{eq:YA}, \ref{eq:YB}, \ref{eq:YC}), the transverse momentum $p^2$ and $\langle p^{2} \rangle_{F}$ only appear in the combined form of the ratio $p^2 / \langle p^{2} \rangle_{F}$. Therefore, in the Bayesian inference, we treat this ratio, denoted as $r = p^2 / \langle p^{2} \rangle_{F}$, as an independent parameter to be inferred. To achieve this, we employ experimental measurements of $c_2\{4\}$, $c_3\{2\}$, $c_3\{4\}$, and $sc_{2,3}\{4\}$  as indirect probes to infer the collective flow dynamics in small systems. These observables are chosen because they are all measured by ATLAS \cite{ATLAS_c24c34,ATLAS_c32,ATLAS_sc23} under the same kinematic cuts, ensuring consistent constraints. 
Specifically, $c_2\{4\}$ primarily constrains $v_2$, 
$c_3\{2\}$ and $c_3\{4\}$ constrain $v_3$, 
and $sc_{2,3}\{4\}$ provides additional information on correlations between $v_2$ and $v_3$, allowing for tighter constraints and more reliable inference of $v_2$ and $v_3$ in small systems. Our approach is grounded in the functional relationships $c_n\{k\} = f(v_2,v_3,r)$ provided by the TMC formalism. These expressions form the basis for constructing the likelihood function, thereby allowing a rigorous Bayesian inference of the posterior probability distributions for the parameters $v_2$, $v_3$, and $r$ from the experimental data.

According to Bayes' theorem, the posterior distribution for the model parameters $\theta$, given the experimental data $y_{\text{exp}}$ with covariance matrix $\Sigma_{\text{exp}}$, is constructed as follows:
\begin{align}
\label{Posterior}
 P(\theta|y_a^{exp}) &= N e^{(-\chi^2/2)} \mathrm{Prior}(\theta) \\
\label{chi2}
 \chi^2 &= \sum_{a,b} (\Sigma_\theta^{-1})_{a,b} \Delta y_a(\theta) \Delta y_b(\theta)
\end{align}
where $\mathcal{N}$ is a normalization constant, and $\Delta y_a(\theta) \equiv y_a(\theta) - y_a^{\text{exp}}$ denotes the residual between the model prediction $y_a(\theta)$ and the experimental measurement $y_a^{\text{exp}}$ for the $a$-th observable. In this work, the covariance matrix $\Sigma$ accounts solely for experimental uncertainties, as the theoretical predictions are computed analytically and are treated as having negligible theoretical error. For the prior $\mathrm{Prior}(\theta)$, we adopt a uniform distribution over the possible parameter range.

To sample the posterior distribution $P(\theta|y_a^{exp})$, we employ the Markov chain Monte Carlo (MCMC) method, which generates representative samples according to the posterior distribution by making a random walk in parameter space weighted by the relative posterior probability. Denoting the parameter set in the $n^{th}$ iteration step as $\theta^{(n)}$, MCMC samples the next step $\theta^{(n+1)}$ as a random walk starting from $\theta^{(n)}$, and decides whether to accept this update according to a probability being $p=min[1,P(\theta^{(n+1)})/P(\theta^{(n)})]$. One can show that samples generated in such a way satisfy the desired posterior distribution $P(\theta|y_a^{exp})$. Markov chains satisfy the property that the conditional probability distribution of future states depends only on the current state and not on the past states preceding the current one, which guarantees the stationary distribution to be achieved.

For the practical computation, we apply the PyMC framework and choose the No-U-Turn Sampler (NUTS), which is the extension to Hamiltonian Monte Carlo method \cite{MCMC2,MCMC1}. We take 50000 samples from each Markov chain with iteration number of adaptive phase being 30000 to ensure decorrelation. With a sufficient amount of parameter sets $(\theta_i)$, we are able to cast them into one- and two-dimensional histograms to measure single-parameter marginal posterior distributions or two-parameter correlations.

\section{Results}
\label{Results}

\begin{figure*}[hbtp]
    \begin{center}
    \includegraphics[trim={3.5cm 3cm 0 3cm}, clip, scale=0.64]{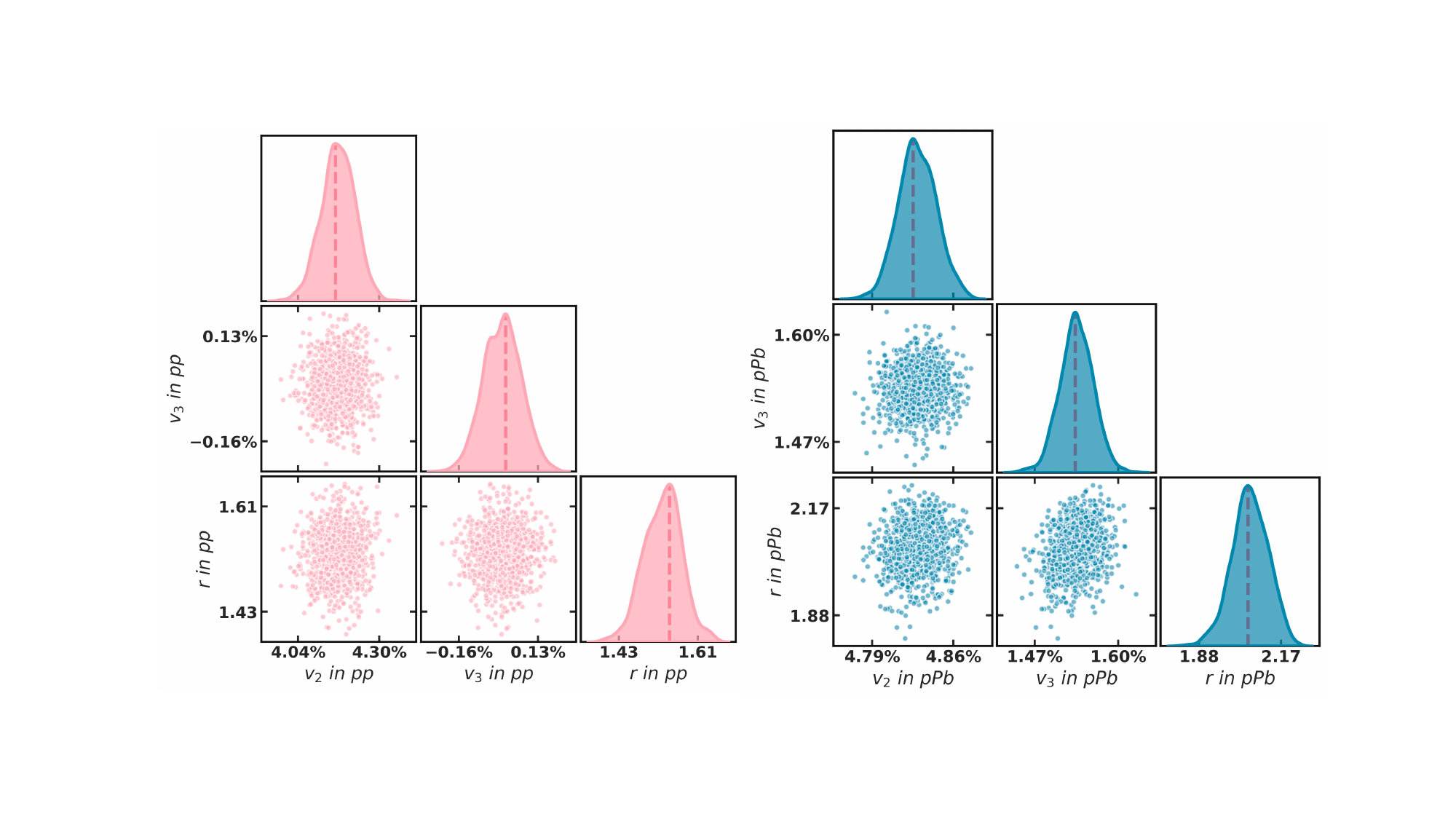}
    \caption{Bayesian posterior probability distributions of $v_2$, $v_3$, 
    and $r$ ($r = p^2 / \langle p^2 \rangle_F$) parameters from MCMC sampling. The left pairplot corresponds to $p$+$p$ collisions, the right to $p$+Pb collisions. Diagonal elements represent marginal posterior distributions via kernel density estimation, and off-diagonal elements show joint posterior distributions as scatter plots, revealing inter-parameter dependencies.}
    \label{posterior}
    \end{center}
\end{figure*}

\label{results}
This section presents the Bayesian inference results to determine the flow harmonics $v_n$ for both $p$+$p$ and $p$+Pb collision systems. The framework is built on calculations from the TMC method and constrained by key experimental observables, namely cumulants and symmetric cumulants, enabling a direct comparison of collective flow behaviors across these two distinct small collision systems. In our Bayesian analysis, we select four experimental observables which are $c_2\{4\}$, $c_3\{2\}$, $c_3\{4\}$, and $sc_{2,3}\{4\}$. The choice of these four observables is motivated by the following two reasons. Firstly, from a physical perspective, they include the information about the collective flow parameters $v_2$ and $v_3$, as well as the correlation between them; secondly, these four observables are the only available experimental data obtained under identical experimental conditions that cover both $p$+$p$ and $p$+Pb collision systems~\cite{ATLAS_sc23,ATLAS_c24c34,ATLAS_c32}. Note that these observables depend on the charged particle multiplicity $N_{\rm ch}$. In our formulas, however, $N$ denotes the number of particles under the influence of the TMC, that is, all particles. Therefore, in the inference we assume a relation between $N$ and $N_{\rm ch}$ based on isospin symmetry, $N = (3/2) N_{\rm ch}$, so that all quantities can be expressed as functions of $N_{\rm ch}$. The validity of this assumption will be discussed in Section~\ref{Sec.results-B}.

Our analysis proceeds in two sequential steps, with consistent methodological application to two collision systems. First, we establish a baseline for each system by treating $v_2$, $v_3$, and $r$ ($r = p^2 / \langle p^{2} \rangle_{F}$) as constants, inferring their values and distributions directly from the measured cumulants $c_2\{4\}$, $c_3\{2\}$, $c_3\{4\}$, and $sc_{2,3}\{4\}$. For the parameters $v_2$, $v_3$, and $r$, we adopt uniform priors within physically motivated ranges based on existing experimental knowledge. In particular, the prior ranges are chosen to fully cover the experimentally observed magnitudes of $v_2$ and $v_3$ in both collision systems, namely $v_2$ in [0.01, 0.06] and $v_3$ in [0, 0.03]. For the parameter $r$, which represents the ratio between the mean squared transverse momentum in the central rapidity region and that over the full phase space, we impose the physically motivated constraint $r>1$.

Second, as the charged-particle multiplicity $N_{\text{ch}}$ increases, the particle density and the frequency of interactions in the system are expected to grow, leading to an enhancement of collective behavior. To systematically investigate the collective flow in small systems, we extend the model by introducing a functional dependence of these parameters on $N_{\text{ch}}$, separately for each collision system. This dependence is parameterized by a power-law form, $v_2 = a \cdot N_{\text{ch}}^b$. Here, the exponent $b$ controls the strength of the power-law scaling and characterizes how rapidly the collective flow evolves with charged-particle multiplicity, while the prefactor $a$ determines the overall strength of collectivity at a reference multiplicity scale. The same parameterization is applied to $v_3$ and $r$ for both $p$+$p$ and $p$+Pb systems. For the exponent $b$, we impose a positive prior, reflecting our assumption that collectivity and transverse momentum do not decrease as $N_{\rm ch}$ increases. The $a$ parameters associated with $v_2$, $v_3$, and $r$ are assigned the same prior range as those adopted in the constant parameter analysis. In addition to the power-law parameterization, we also explored alternative functional forms motivated by prior expectations of increasing collectivity with multiplicity, such as $v_2 = v_0 \cdot \arctan(S\cdot N_{\text{ch}})$ and polynomial parameterizations. However, in these cases, the posterior distributions exhibited poor convergence behavior and unstable parameter correlations. We therefore adopted the power-law form because of its relative simplicity, clearer physical interpretation of the parameters, and significantly improved stability of the resulting MCMC posterior distributions.

\subsection{Bayesian Inference of Constant Parameters}

\begin{figure*}[hbtp]
    \begin{center}
	\includegraphics[scale=0.2]{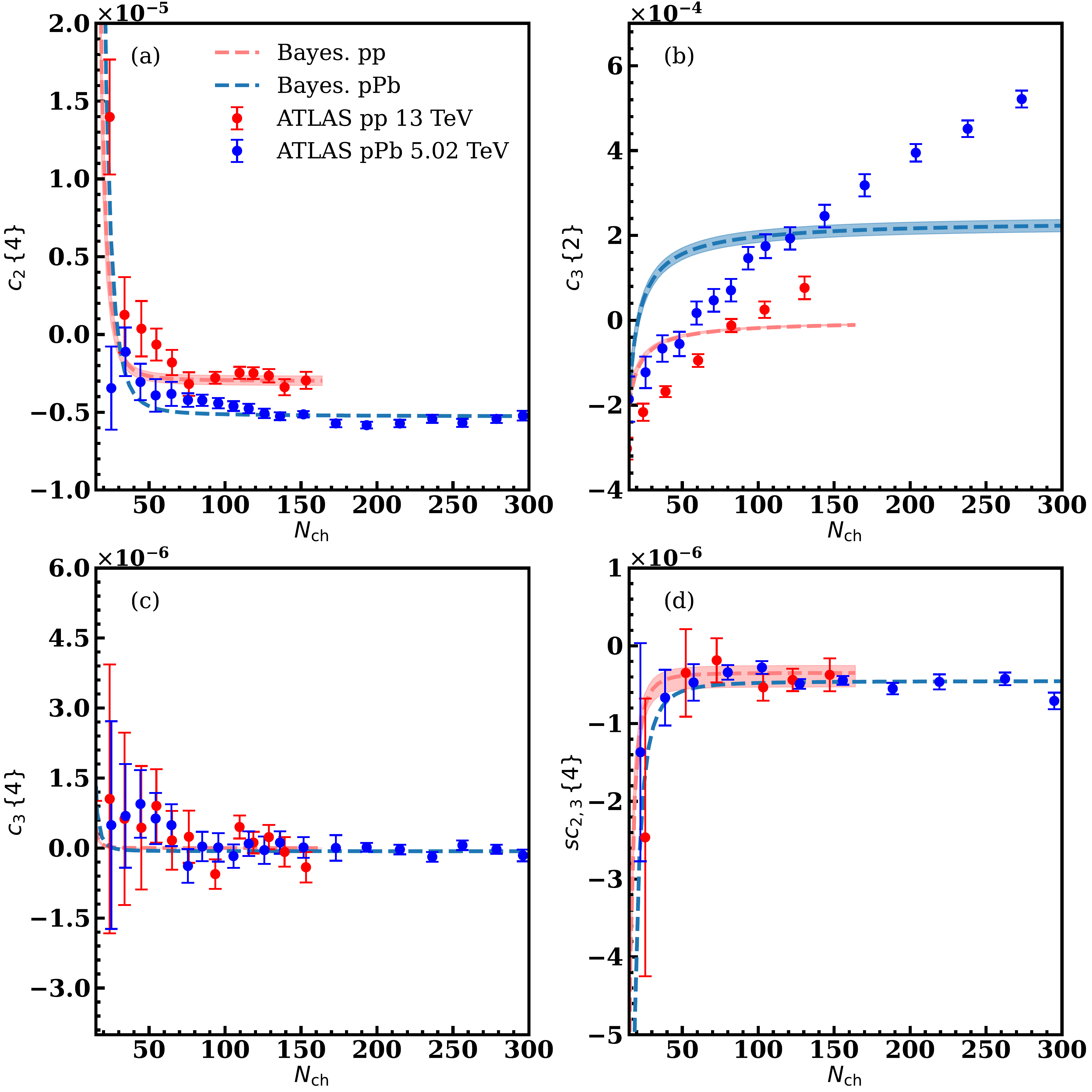}
	\caption{Comparison between Bayesian inference results and ATLAS experimental data for (a) $c_{2}\{4\}$, (b) $c_{3}\{2\}$, (c) $c_{3}\{4\}$, and (d) $sc_{2,3}\{4\}$. Data points represent experimental measurements, while curves show analytical results obtained by substituting Bayesian-inferred parameters into the TMC formula. Red symbols and curves correspond to the $p$+$p$ system, and blue to the $p$+Pb system.}
	\label{constant_cumulants}
    \end{center}
\end{figure*}

The posterior distributions of the parameters obtained from the Markov Chain Monte Carlo (MCMC) sampling are shown in Fig.~\ref{posterior}, which displays pairwise parameter correlations and marginalized probability densities for both $p$+$p$ and $p$+Pb collision systems. Multiple independent Markov chains were tested and found to converge consistently. Furthermore, the posterior distributions are well constrained and significantly narrower than the corresponding prior ranges, indicating that the inferred results are primarily driven by the experimental data rather than by prior assumptions. For $p$+$p$  collisions, the maximum a posteriori (MAP) parameter values of the key baseline parameters are: $v_2^{\mathrm{pp, MAP}} = 0.0416^{\,+\,0.0005}_{\,-\,0.0006}$, 
$v_3^{\mathrm{pp, MAP}} = 0.0001^{\,+\,0.0007}_{\,-\,0.0007}$, and 
$r^{\mathrm{pp, MAP}} = 1.5434^{\,+\,0.0240}_{\,-\,0.0569}$~GeV/$c$; on the other hand, the $p$+Pb  system exhibits higher maximum probability values at the baseline: $v_2^{\mathrm{pPb, MAP}} = 0.0482^{\,+\,0.0002}_{\,-\,0.0001}$, 
$v_3^{\mathrm{pPb, MAP}} = 0.0153^{\,+\,0.0003}_{\,-\,0.0002}$, and 
$r^{\mathrm{pPb, MAP}} = 2.0559^{\,+\,0.0740}_{\,-\,0.0554}$~GeV/$c$. These values, representing the most probable parameter estimates from the MCMC posterior, are subsequently used as core inputs for calculating the cumulants, which are then compared with experimental data.

After obtaining the posterior distributions of the inferred parameters, the cumulants are evaluated for each posterior sample. The predictions corresponding to MAP parameter values are taken as the central estimates. The associated uncertainties are quantified by the 95\% credible intervals of the cumulant distributions constructed from the posterior samples. The resulting cumulants and symmetric cumulants for both $p$+$p$ and $p$+Pb systems are then compared with the experimental data, as shown in Fig.~\ref{constant_cumulants}. Depending on the maximum measured charged-particle multiplicity, our model results constrained by the experimental data are shown up to $N_{\mathrm{ch}}=160$ for the $p$+$p$ system, but up to $N_{\mathrm{ch}}=300$ for the $p$+Pb system. The same presentation scheme is adopted in the subsequent figures for consistency. For the four-particle cumulant $c_3\{4\}$, Fig.~\ref{constant_cumulants} (c), and the symmetric cumulant $sc_{23}\{4\}$, Fig.~\ref{constant_cumulants} (d), the TMC-calculated values, with error bars reflecting the propagated uncertainties, show good agreement with the experimental data over the entire multiplicity range in both $p$+$p$ and $p$+Pb collision systems within large experimental error bars. These results demonstrate the feasibility of extracting collective flow in small systems using the Bayesian inference approach.

However, distinct discrepancies arise for $c_2\{4\}$ in Fig.~\ref{constant_cumulants} (a) and $c_3{\{2\}}$ in Fig.~\ref{constant_cumulants} (b), especially for $c_3{\{2\}}$. The experimental data for $c_3{\{2\}}$ exhibits a clear upward trend, with increasing charged-particle multiplicity $N_\text{ch}$, while the calculated $c_3{\{2\}}$ values remain nearly constant at large $N_\text{ch}$ due to the constant parameters in the baseline model. This mismatch directly demonstrates that treating $v_2$, $v_3$, and $r$ as constants is insufficient to capture the real dependence of $c_2\{4\}$ and $c_3\{2\}$ observed in experiments, highlighting the need for an extended model with multiplicity-dependent parameters to capture the dependence of collectivity on $N_\text{ch}$.

\subsection{Bayesian Inference of $N_{\text{ch}}$-Dependent Parameters}
\label{Sec.results-B}
\begin{figure*}[hbtp]
    \begin{center}
	\includegraphics[scale=0.2]{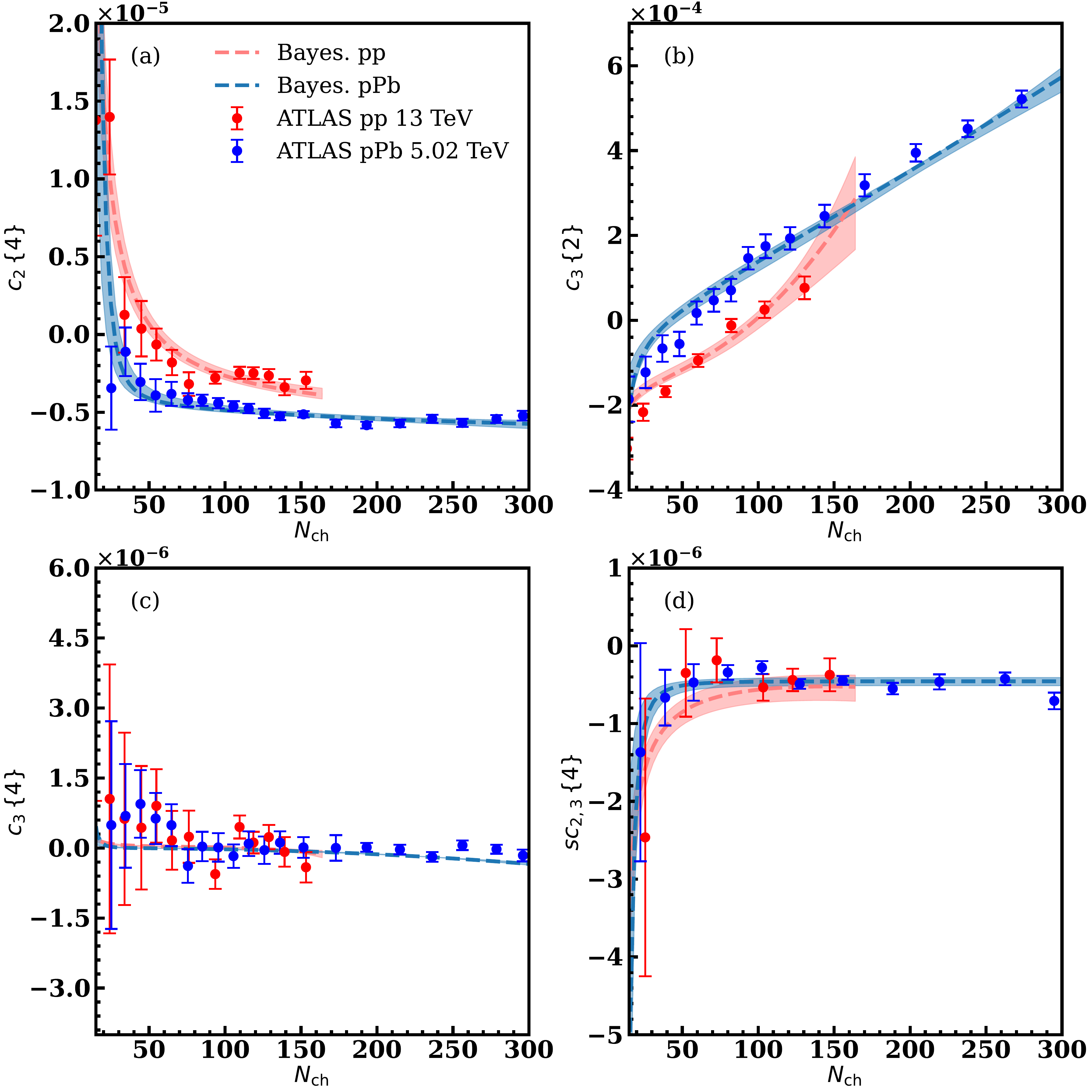}
	\caption{Comparison between the Bayesian inference results with ATLAS experimental measurements for (a) $c_{2}\{4\}$, (b) $c_{3}\{2\}$, (c) $c_{3}\{4\}$, and (d) $sc_{2,3}\{4\}$. Symbols denote experimental measurements, and curves show TMC results using Bayesian-inferred parameters that follow a power-law scaling with $N_\text{ch}$. Red (blue) corresponds to $p$+$p$ ($p$+Pb) collisions.}
	\label{dependence_cumulants}
    \end{center}
\end{figure*}

\begin{figure*}[hbtp]
    \begin{center}
    \includegraphics[scale=0.18]{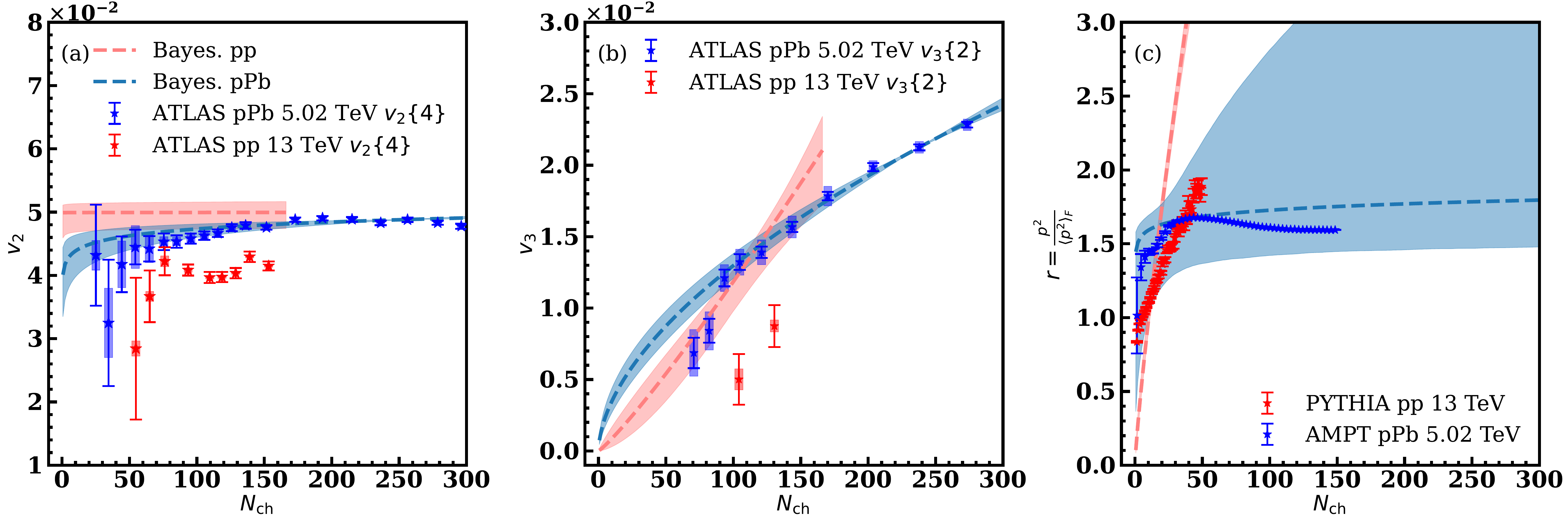}
	\caption{Comparison of (a) $v_2$ and (b) $v_3$ values extracted from cumulants with experimental data, and the parameter (c) $r$ with model calculations, using PYTHIA 8 for $p$+$p$ and AMPT for $p$+Pb. Red and blue symbols/lines represent the $p$+$p$ and $p$+Pb systems, respectively.}
	\label{dependence_parameter}
    \end{center}
\end{figure*}

To investigate the dependence of collective flow on $N_\text{ch}$, we adopt a power-law parameterization, e.g.,  $v_2 = a \cdot N_{\text{ch}}^b$ , which provides a simple yet flexible form to describe the observed multiplicity dependence of collective flow. After obtaining the posterior distributions of the parameters in the power-law form for $v_2$, $v_3$, and $r$, we extract the corresponding flow coefficients and transverse momentum ratio, and determine their dependence on the charged-particle multiplicity $N_{\text{ch}}$ in small collision systems. We further verified that the posterior distributions of all parameters are well converged, with the Gelman–Rubin convergence statistic R-hat for all parameters approaching 1. This indicates good consistency among independent Markov chains and demonstrates the stability and reliability of the MCMC sampling procedure. These are subsequently incorporated into the TMC formalism, with a full propagation of the parameter uncertainties from the posterior distributions. As illustrated in Fig.~\ref{dependence_cumulants}, this approach leads to an improved description of the measured cumulants, which is further confirmed by significantly smaller $\chi^2$ values compared to the constant-parameter model discussed in the previous subsection.

Through Bayesian inference within the TMC framework constrained by experimental cumulant data, we extract the $N_{\text{ch}}$-dependent collective flow, free from TMC contamination. In particular, Fig.~\ref{dependence_parameter} (a) compares our TMC-inferred results of $v_2$ with experimental measurements of $v_2\{4\}$ in $p$+$p$ and $p$+Pb collisions. We observe that the extracted TMC-free collective flow coefficient $v_2$ is generally similar between $p$+$p$ and $p$+Pb collision systems. In particular, for the $p$+$p$ system, the inferred power-law parameter $b$ is found to be consistent with zero, indicating that $v_2$ exhibits little to no dependence on $N_{\text{ch}}$ and can be regarded as approximately constant. At low multiplicities, however, $v_2$ in $p$+$p$ appears slightly larger than in $p$+Pb, which may be attributed to stronger non-flow contributions beyond TMC in $p$+$p$ collisions. This overall similarity is expected, if the mechanism responsible for the generation of collective flow in small systems is the same in both cases, namely arising from event-by-event fluctuations of energy density in the initial state. In $p$+Pb collisions, our inferred TMC-free collective flow $v_2$ is in excellent agreement with the experimental measurements of $v_2\{4\}$, which are directly obtained from $v_2\{4\}=\sqrt[4]{-c_2\{4\}}$. Note that the $v_n$ values, along with uncertainty bands, are obtained from Bayesian analysis under the TMC framework, reflecting the flow after subtracting TMC contributions. This agreement suggests that, in $p$+Pb systems, the non-flow background contribution from TMC is effectively suppressed in the four-particle azimuthal correlations with the sub-event cumulant method.

In Fig.~\ref{dependence_parameter} (b), we compare our TMC-inferred results of extracted $v_3$ with experimental measurements of $v_3\{2\}$ in $p$+$p$ and $p$+Pb collisions. Note that, in both two collision systems, the two-particle cumulant $c_{3}\{2\}$ becomes negative at low multiplicities, such that $v_{3}\{2\}=\sqrt{c_{3}\{2\}}$ is not defined in this region. Consequently, the experimental measurements of $v_{3}\{2\}$ are available only at higher multiplicities, resulting in fewer data points compared to those for $c_{3}\{2\}$. We observe that $v_3$ in $p$+$p$ is smaller than that in $p$+Pb at low multiplicities. However, with increasing $N_{\mathrm{ch}}$, the $v_3$ values between the two systems become consistent within the uncertainties. For $v_3$ in $p$+Pb collisions, our results exhibit excellent agreement with experimental measurements at high $N_{\text{ch}}$, while at low $N_{\text{ch}}$ they are slightly higher than the values extracted from experimental cumulants. For $p$+$p$ collisions, the extracted $v_2$ and $v_3$ values are systematically higher than the experimental measurements over the entire $N_{\text{ch}}$ range. We will interpret these differences by examining the relative contributions of different components to the measured cumulants. 

The extracted parameter $r = p^2 / \langle p^2 \rangle_F$ is presented in Fig.~\ref{dependence_parameter} (c). At low $N_{\text{ch}}$, the value of $r$ in $p$+Pb collisions is larger than that in $p$+$p$ collisions; As $N_{\text{ch}}$ increases, $r$ in $p$+Pb collisions saturates, whereas in $p$+$p$ collisions it continues to increase and exceeds that in $p$+Pb. This trend and ordering are consistent with the experimentally observed behavior of mean transverse momentum in the two systems~\cite{meanpT}. To further validate our extraction, we compare the inferred $r = p^2 / \langle p^2 \rangle_F$ with the PYTHIA model~\cite{PYTHIA_k} calculation in $p$+$p$ collisions and the AMPT model~\cite{AMPT_k} calculation in $p$+Pb collisions, where $p^2$ is taken as the mean squared transverse momentum in the central rapidity region and $\langle p^2 \rangle_F$ is evaluated over the full phase space. The model results reproduce the relative magnitude and trend of the extracted values of $r$ in both $p$+$p$ and $p$+Pb collision systems. 

In principle, $N$, the number of particles subject to TMC, could be treated as an additional free parameter. However, as seen, e.g., in Eq.~(\ref{eq:YA}), it always appears together with $r$; consequently, $N$ and $r$ cannot be treated as independent parameters. In our approach, we assume the relation between $N$ and $N_{\text{ch}}$, $N = (3/2)N_{\text{ch}}$, and determine $r$. The obtained value of $r$ is consistent with model calculations, supporting the validity of this relation between $N$ and $N_{\text{ch}}$. Since N and $r$ always appear in the combined form $r/N$ within the TMC framework, the dependence of N on $N_{\rm ch}$ can be effectively absorbed into the parameterization of $r$, and therefore does not affect the extracted values of $v_2$ and $v_3$. To verify this explicitly, we have tested the same analysis using the alternative assumption $N=3 N_{\rm ch}$, and found that the extracted $v_2$ and $v_3$ remain essentially unchanged, while the extracted values of $r$ increase approximately by a factor of two, as expected.

\begin{figure*}[hbtp]
    \begin{center}
	\includegraphics[scale=0.2]{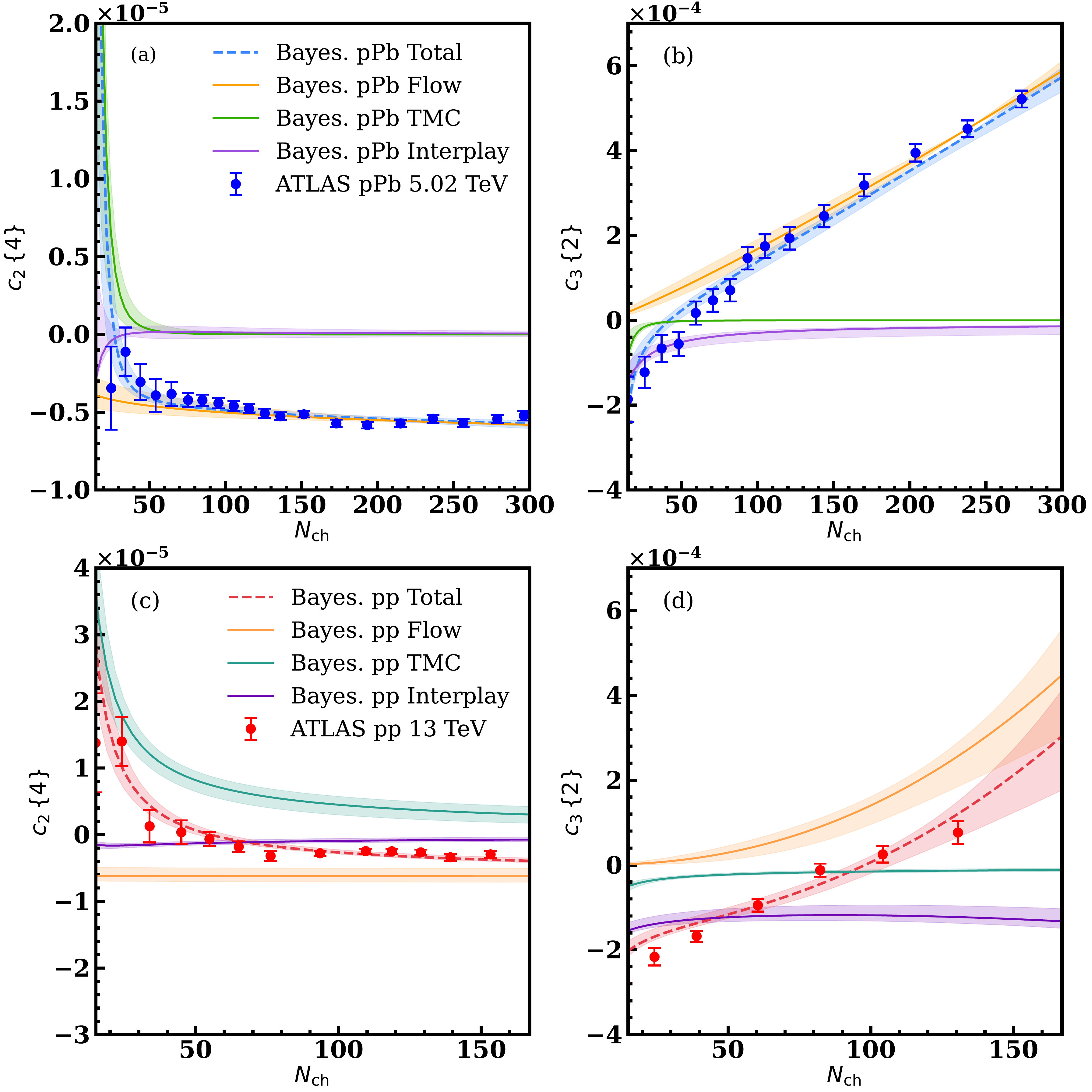}
	\caption{Comparison between the Bayesian analytical results and experimental data for (a) $c_{2}\{4\}$ and (b) $c_{3}\{2\}$ in $p$+Pb collisions, and (c) $c_{2}\{4\}$ and (d) $c_{3}\{2\}$ in $p$+$p$ collisions, decomposed into different physical contributions.  The yellow, green, and purple dashed lines represent the contributions from pure flow, pure TMC, and their interplay, respectively.}
	\label{cumulants_flow_tmc_interplay}
    \end{center}
\end{figure*}

\subsection{Comparative decomposition of physical contributions of Bayesian-inferred results}

To understand the differences between the Bayesian-inferred $v_n$ and the experimental measurements in $p$+$p$ and $p$+Pb collisions, we decompose the theoretically calculated cumulants into three distinct contributions. This decomposition allows a clear interpretation of the experimental results, as illustrated in Fig.~\ref{cumulants_flow_tmc_interplay}. The upper panels show $c_2\{4\}$ and $c_3\{2\}$ in $p$+Pb collisions, while the lower two subfigures correspond to the $p$+$p$ system. As introduced in Section~\ref{TMC}, we analyze three distinct contributions, including the collective flow term, the TMC-only term, and the interplay term between TMC and flow, which are depicted as solid curves in yellow, green, and purple, respectively.

We first discuss the $p$+Pb system. For $c_2\{4\}$, as shown in Fig.~\ref{cumulants_flow_tmc_interplay} (a), the contribution from pure collective flow alone (indicated by the yellow band) nearly reproduces all experimental data points when the multiplicity $N_{\mathrm{ch}}$ is larger than 50. 
For  $c_{3}\{2\}$ in Fig.~\ref{cumulants_flow_tmc_interplay} (b), the negative values at low $N_{\mathrm{ch}}$ arise primarily from the interplay between TMC and collective flow, clearly indicated by the purple band. As the multiplicity increases, the contribution from pure flow grows and eventually dominates at higher $N_{\mathrm{ch}}$, providing a good match with experimental data. Consequently, the extracted $v_3$ (Fig.~\ref{dependence_parameter}) is slightly higher than measurements at low $N_{\mathrm{ch}}$, consistent with the yellow curve lying above the data in this region. This indicates that in $p$+Pb, non-flow contributions to $v_3$ at low multiplicities are mainly due to TMC, which our framework quantitatively captures.

Turning to $p$+$p$ collisions, for $c_{2}\{4\}$ in Fig.~\ref{cumulants_flow_tmc_interplay} (c), the magnitude of the pure flow contribution (yellow band) exceeds that of negative experimental points, explaining why the extracted $v_2$ (Fig.~\ref{dependence_parameter}) is systematically higher. In $p$+$p$ collisions, the TMC contribution cannot be neglected at any $N_{\mathrm{ch}}$, as evidenced by the consistently nonzero green band. Our Bayesian approach enables the separation and subtraction of TMC-induced non-flow, allowing the extraction of the TMC-free collective flow $v_2$. 
For $c_{3}\{2\}$ in $p$+$p$ collisions, as shown in Fig.~\ref{cumulants_flow_tmc_interplay} (d), the pure flow contribution is actually larger than the direct reflection of the experimental data, because the interplay term is significant across all $N_{\mathrm{ch}}$, resulting in a negative contribution to the experimental data. It should be noted that at low $N_{\mathrm{ch}}$, our result cannot fully explain the experimental results, which suggests that, besides TMC, other unknown effects may also contribute to $c_{3}\{2\}$ in $p$+$p$ collisions.

\begin{figure*}[hbtp]
    \begin{center}
	\includegraphics[scale=0.2]{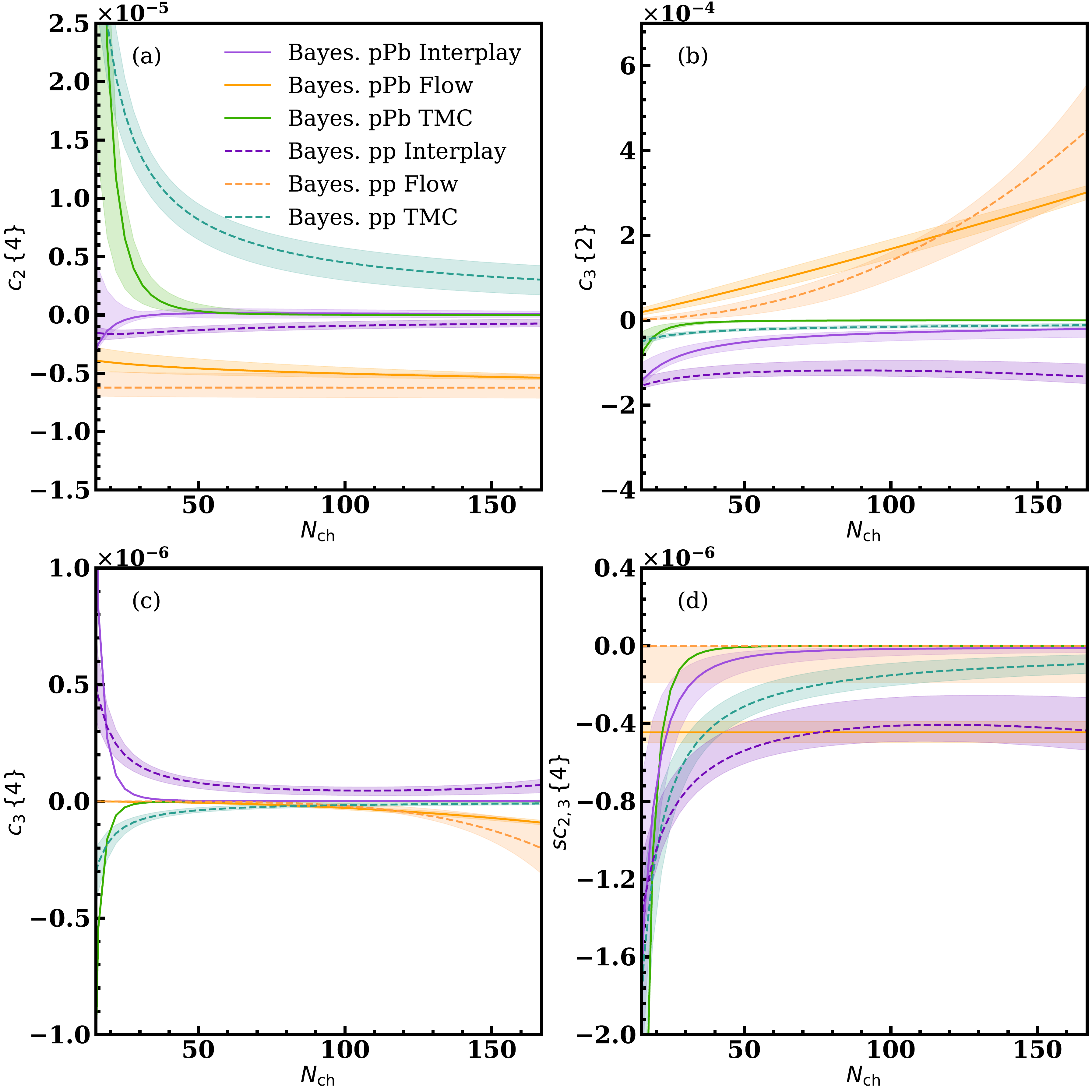}
	\caption{Comparison of different physical contributions to (a) $c_2\{4\}$, (b) $c_3\{2\}$, (c) $c_3\{4\}$ and (d) $sc_{2,3}\{4\}$ between $p$+$p$ and $p$+Pb systems. The dashed lines correspond to the $p$+$p$ system, and the solid lines to the $p$+Pb system. Colors indicate the contributions: yellow for pure flow, green for pure TMC, and purple for the interplay between flow and TMC.}
	\label{cumulants_flow_tmc_interplay_comparetwosys}
    \end{center}
\end{figure*}

Finally, we perform a detailed comparison of the individual contributions to different cumulants between the two collision systems. We observe that, for $c_{2}\{4\}$ in Fig.~\ref{cumulants_flow_tmc_interplay_comparetwosys} (a), the pure flow contribution shows no significant difference between $p$+$p$ and $p$+Pb systems within uncertainties, while a stronger TMC effect is observed in the $p$+$p$ system than in $p$+Pb systems. For $c_{3}\{2\}$ in Fig.~\ref{cumulants_flow_tmc_interplay_comparetwosys} (b), a pronounced difference is observed in the interplay term between the two systems, indicating distinct correlations between flow and the TMC effect. Furthermore, in $\mathrm{sc}_{23}\{4\}$ in Fig.~\ref{cumulants_flow_tmc_interplay_comparetwosys} (d), the visible difference in the yellow bands clearly demonstrates a markedly different correlation between $v_{2}$ and $v_{3}$ in $p$+$p$ and $p$+Pb collisions. Specifically, the $v_{2}$--$v_{3}$ correlation in $p$+$p$ collisions is found to be consistent with zero, whereas a significantly negative correlation is observed in the $p$+Pb system. These observations indicate that, although the extracted $v_{2}$ and $v_{3}$ values are similar in the two small collision systems, the associated TMC effect, the interplay between TMC and flow, and the $v_{2}$--$v_{3}$ correlation differ significantly between $p$+$p$ and $p$+Pb systems. The difference in the interplay contribution between TMC and collective flow is mainly driven by the different TMC contributions in the two collision systems, since the extracted flow components themselves are found to be similar. Within the present framework, the TMC contribution depends primarily on the particle number $N$ and the transverse momentum $p$ of produced particles. Therefore, differences in particle spectra, transverse momentum distributions, multiplicity fluctuations, acceptance effects, and collision energies between $p$+$p$ and $p$+Pb systems can all affect the magnitude of the TMC contribution, and consequently lead to different interplay effects between TMC and collective flow. Our Bayesian analysis based on the TMC framework offers a novel tool for uncovering the physical source distinctions hidden within similar experimental data across different small collision systems.

\section{Discussions}

For the two-particle cumulant $c_{2}\{2\}$, which is not utilized in this work, the main limitation is that it contains significant nonflow contributions beyond the TMC effect in small systems, which are not removed or sufficiently suppressed. Therefore, the present framework based solely on TMC and flow contributions is not able to fully describe the experimental $c_{2}\{2\}$ data~\cite{c22_exp,c22}. Nevertheless, within the framework developed here, $c_{2}\{2\}$ may provide a potential way to estimate additional nonflow contributions beyond the TMC effect, such as jet-related correlations, by comparing the measured $c_{2}\{2\}$ with the calculated TMC and flow components. However, beyond the observables investigated in this work, the proposed framework can be further extended to a broader range of flow measurements and collision systems. As higher-order flow observables become available, this approach provides a systematic way to isolate flow from TMC effect. For instance, by incorporating higher-order cumulants such as $c_4\{2\}$ and symmetric cumulants such as $sc_{2,4}\{4\}$, the framework can be applied to constrain higher-order flow harmonics, including $v_4$. Such extensions would enable a more comprehensive characterization of the system-size dependence and the possible origins of collective behavior in small systems.

Moreover, the methodology developed here is not limited to the ATLAS measurements considered in this study, but can be applied to multi-particle correlation measurements from different experimental programs and collision systems. Existing measurements of multi-particle cumulants and symmetric cumulants from experiments such as ALICE and CMS also provide promising opportunities for performing TMC-free extraction of collectivity over a much wider range of systems. For ALICE, multi-particle cumulants, including $c_3\{2\}$, $c_2\{4\}$, as well as symmetric cumulants such as $sc(3,2)$ and $sc(4,2)$, have been measured systematically in various collision systems, including $p$+$p$, $p$+Pb, Xe+Xe, and Pb+Pb collisions \cite{LHC6}. Applying the present framework to these observables would enable a consistent comparison of the collective component after removing TMC effects, providing further insight into the evolution of collectivity with system size. For CMS, higher-order cumulants such as $c_2\{4\}$, $c_2\{6\}$, and $c_2\{8\}$ have been measured in small collision systems \cite{CMS_impact}. Incorporating these observables into the present framework would allow the extraction of a TMC-free $v_2$ signal and provide a more stringent test of the collective origin of anisotropic flow in small systems through comparisons among different cumulant orders. Furthermore, we have performed a preliminary investigation of the applicability of this framework to forward-rapidity measurements, where the detector acceptance is highly asymmetric, such as those performed by LHCb. Using PYTHIA simulations within the relevant forward-rapidity acceptance, we selected events satisfying the transverse momentum conservation constraint and calculated the corresponding multi-particle cumulants and found that the cumulants obtained from the TMC-constrained event sample show good agreement with the theoretical expectation, indicating that the proposed framework may also be applicable in the forward-rapidity region. These developments will help establish a unified interpretation of flow observables across different experiments and provide further constraints on the nature of collectivity in small collision systems.

It is also important to note that the present framework involves several approximations and assumptions, which may introduce potential uncertainties. The first approximation originates from the use of the Central Limit Theorem in deriving the analytical expression of the TMC contribution. This approximation becomes increasingly accurate with increasing particle multiplicity. Our Monte Carlo studies demonstrate that, within the multiplicity range relevant to the experimental data analyzed in this work, the analytical TMC calculation agrees well with the numerical results, indicating that the uncertainty associated with this approximation remains small over most of the analyzed multiplicity range, while it may become more relevant in the lowest-multiplicity bins.
The second approximation is related to the truncation of the expansion used in calculating the multi-particle cumulants. We retain the leading non-vanishing TMC terms and neglect higher-order corrections. By explicitly evaluating the higher-order contributions, we find that their effects are only relevant at very low multiplicity, while they become negligible for the multiplicity range considered in the present analysis. Therefore, the uncertainty introduced by this approximation is expected to have a limited influence on the final results.
Thirdly, we emphasize that the present framework assumes TMC to be the dominant long-range nonflow contribution and does not explicitly include other possible sources of nonflow, such as jet correlations and resonance decays. Although multi-particle cumulants, especially higher-order cumulants, are generally effective in suppressing nonflow contributions, residual effects may still remain. Therefore, the TMC-free observables extracted in this work should be interpreted as the collective component after removing the dominant TMC-induced correlation, rather than a completely model-independent measurement of collectivity. A more comprehensive treatment including additional nonflow mechanisms will be an important direction for future developments of this framework.
In addition, in the analytical calculations presented in Eqs.~(\ref{eq:c24}, \ref{eq:c34}, \ref{eq:sc23}), event-by-event fluctuations are not explicitly considered. Therefore, the extracted parameters should be regarded as approximations of their corresponding mean values. In a more complete treatment, fluctuations of the flow coefficients, including $v_2$ , $v_3$, $p$, as well as their correlation coefficient $\rho_n \left ( v_n\{2\}^{2} , p \right ) $, should be incorporated. The inclusion of these event-by-event fluctuations will be an important extension of the present framework and will be pursued in future studies.

\section{Conclusions}
\label{summary}

This work employs a Bayesian inference framework, built upon the transverse momentum conservation (TMC) formalism and constrained by the experimental measurements of $c_2\{4\}$, $c_3\{2\}$, $c_3\{4\}$, and $sc_{2,3}\{4\}$, to extract the TMC-free collective flow behavior in $p$+$p$ and $p$+Pb collisions. The analysis is conducted in two stages: first, a baseline model with constant parameters is established for each collision system; then, a refined model introduces a power-law dependence of $v_2$, $v_3$, and $r$ ($r = p^2 / \langle p^{2} \rangle_{F}$)  on charged-particle multiplicity. The power-law model significantly improves the description of the experimental cumulant data, yielding substantially lower $\chi^2$ values compared to the constant-flow model. 

Within the TMC theoretical framework, our approach enables a systematic disentanglement of TMC-induced nonflow effects, allowing the extraction of the collective component $v_n$ remaining after accounting for the modeled TMC contribution. This capability is essential for studying the nature of collectivity in small collision systems, where nonflow contributions are known to be significant. We observe that the extracted collective flow coefficients $v_2$ and $v_3$ are similar between $p$+$p$ and $p$+Pb collision systems. This observation indicates that, after removing the modeled TMC contributions, the remaining TMC-free collective flow signals in the two systems are largely consistent with each other. However, the contributions from TMC, the interplay between TMC and collective flow, as well as the correlations between $v_2$ and $v_3$, exhibit clear system-dependent differences. This demonstrates that, beyond the similar TMC-free collectivity, more detailed and system-specific correlations differ between $p$+$p$ and $p$+Pb collisions.

Moreover, this framework directly exposes the impact of TMC on flow observables and provides a quantitative assessment of flow  with the nonflow contribution from TMC effects removed in small systems. In $p$+Pb collisions, the measurement of $c_2\{4\}$ effectively suppresses nonflow effects, in particular the TMC contributions, such that the extracted $v_2\{4\}$ is dominated by TMC-free collective flow. In contrast, for $c_3\{2\}$, the interplay between TMC and flow leads to a negative contribution, which dominates at low charged-particle multiplicity. Consequently, the experimentally measured $c_3\{2\}$ becomes negative in this region, and the extracted TMC-free $v_3$ is larger than the corresponding experimental $v_3\{2\}$. With increasing multiplicity, the TMC-free flow contribution gradually becomes dominant, causing the extracted $v_3$ to approach the experimental measurements. In $p$+$p$ collisions, however, due to the fact that TMC effects remain substantial across the entire multiplicity range, and contribute positively to $c_2\{4\}$ while negatively to $c_3\{2\}$, the experimentally measured $v_2\{4\}$ and $v_3\{2\}$ are systematically suppressed relative to the extracted TMC-free collectivity.

Overall, the combination of the TMC framework with hierarchical Bayesian inference offers a powerful approach to studying the nature of collective flow via azimuthal correlations in small collision systems. This methodology has the potential to establish a new paradigm for isolating collectivity from TMC-induced nonflow effects, laying a useful foundation for a clear understanding of collective phenomena in small collision systems.

\begin{acknowledgments}

Shuang Guo thanks Prof. Kai Zhou for his introductory instruction in Bayesian inference framework. This work is partially supported by the National Natural Science Foundation of China under Grants  No. 12325507, No. 12547102, and No. 12147101, the National Key Research and Development Program of China under Grant No. 2022YFA1604900 (S.G., J.P. and G.M.), the Ministry of Science and Higher Education (PL), and the National Science Centre (PL), Grant No. 2023/51/B/ST2/01625 (A.B.).\\
\\
\textbf{Conflict of Interest} The authors declare that they have no conflict of interest.

\end{acknowledgments}

\bibliographystyle{plainnat}
\bibliography{ref}
\end{CJK*}
\end{document}